\RequirePackage[2022-06-01]{latexrelease}
\documentclass[aps,prl,10pt,twocolumn,preprintnumbers,superscriptaddress,showpacs,floatfix,nofootinbib,longbibliography]{revtex4-2}
\usepackage{textcomp,gensymb}
\usepackage{hyperref}
\usepackage{graphicx}
\usepackage{amsfonts,amsmath,amssymb,bm,bbm}
\usepackage{color,soul}
\usepackage[dvipsnames]{xcolor}
\usepackage{slashed}
\usepackage{flushend}
\usepackage{physics}
\usepackage{graphicx}
\usepackage{dcolumn}
\usepackage{float}
\usepackage{ragged2e}
\usepackage{subcaption}
\usepackage{caption}
\usepackage{cleveref}

\hypersetup{
    pdfnewwindow=true,      
    colorlinks=true,       
    linkcolor=blue,          
    citecolor=blue,        
    filecolor=blue,      
    urlcolor=blue        
}

\begin{document}
\preprint{IPPP/26/50}
\title{Charged Lepton Flavor Violation at Neutrino Telescopes}

\author{Writasree Maitra}
\email{m.writasree@wustl.edu}
\affiliation{Department of Physics, Washington University, St. Louis, MO 63130, USA}

\author{Carlos A. Arg\"uelles}
\email{carguelles@fas.harvard.edu}
\affiliation{Department of Physics \& Laboratory for Particle Physics and Cosmology, Harvard University, Cambridge, MA 02138, USA}

\author{P. S. Bhupal Dev}
\email{bdev@wustl.edu}
\affiliation{Department of Physics, Washington University, St. Louis, MO 63130, USA}
\affiliation{McDonnell Center for the Space Sciences, Washington University, St. Louis, MO 63130, USA}
\affiliation{PRISMA$^{++}$ Cluster of Excellence \& Mainz Institute for Theoretical Physics, 
Johannes Gutenberg-Universit\"{a}t Mainz, 55099 Mainz, Germany}

\author{Ivan Martinez-Soler}
\email{ivan.j.martinez-soler@durham.ac.uk}
\affiliation{Institute for Particle Physics Phenomenology, Durham University, South Road DH1 3LE, Durham, U.K.}

\author{Manibrata Sen}
\email{manibrata@iitb.ac.in}
\affiliation{Department of Physics, Indian Institute of Technology Bombay, Powai, Maharashtra 400076, India}

\begin{abstract}
Any observation of charged lepton flavor violation (CLFV) would be a clear signal of beyond-the-Standard-Model physics. 
Here, we propose a novel CLFV search using neutrino telescopes with their large cosmic-ray muon samples. 
Specifically, we use a recent IceCube cosmic-ray muon dataset and propose a new search for muon-to-tau conversion inside the IceCube detector.
We illustrate our idea with  CLFV interactions described by model-independent Effective Field Theory (EFT) operators and present the IceCube sensitivity on the relevant EFT scale. 
We also consider a specific realization of the EFT operator in terms of an axial-vector $Z'$ interaction and show sensitivities in the $Z'$ mass-coupling plane. 
We compare our sensitivities with those from low-energy CLFV searches, as well as from current and future collider experiments.
We also show projections from next-generation neutrino telescopes, such as IceCube-Gen2 and HUNT, and demonstrate how neutrino telescopes can provide a powerful complementary probe of CLFV.
\end{abstract}
\maketitle
{\textbf {Introduction.--}} 
The observation of neutrino oscillations between all three flavors indicates that lepton flavor is not conserved. It also motivates an extension of the Standard Model (SM) that incorporates neutrino mass, which inevitably gives rise to lepton flavor violation (LFV) in the charged lepton sector as well. Within the minimal SM extension with nonzero neutrino mass, however, the charged LFV (CLFV) processes are extremely suppressed by the neutrino mass~\cite{Petcov:1976ff}. This makes experimental searches for CLFV particularly compelling, since any measurable signal would point to physics beyond the minimally extended SM. In general, many beyond-the-SM (BSM) scenarios introduce new sources of CLFV, potentially accessible at current and future experiments~\cite{deGouvea:2013zba,Bernstein:2013hba, Calibbi:2017uvl,Ardu:2022sbt,Davidson:2022jai}. Any positive  observation of CLFV would be an unambiguous sign of BSM physics. It might also  provide insight into the origin of neutrino mass, and even the matter-antimatter asymmetry of the Universe via (flavored) leptogenesis~\cite{Fukugita:1986hr,Dev:2017trv}.

\par Over the past decade, Cherenkov neutrino telescopes have opened a new observational window into the high-energy Universe. 
Foremost among these experiments is the IceCube detector at the South Pole~\cite{IceCube:2016zyt}, which has detected numerous astrophysical neutrino events~\cite{IceCube:2013cdw,IceCube:2013low,IceCube:2014stg,IceCube:2021rpz,IceCube:2022der,IceCube:2023ame}.
The power of Cherenkov neutrino telescopes stems from their ability to identify high-energy charged leptons produced in neutrino–nucleon interactions.
In this sense, they function as charged-lepton detectors, opening a novel and largely unexplored avenue for CLFV searches.

\begin{figure}[!t]
    \centering
    \includegraphics[width=0.8\linewidth]{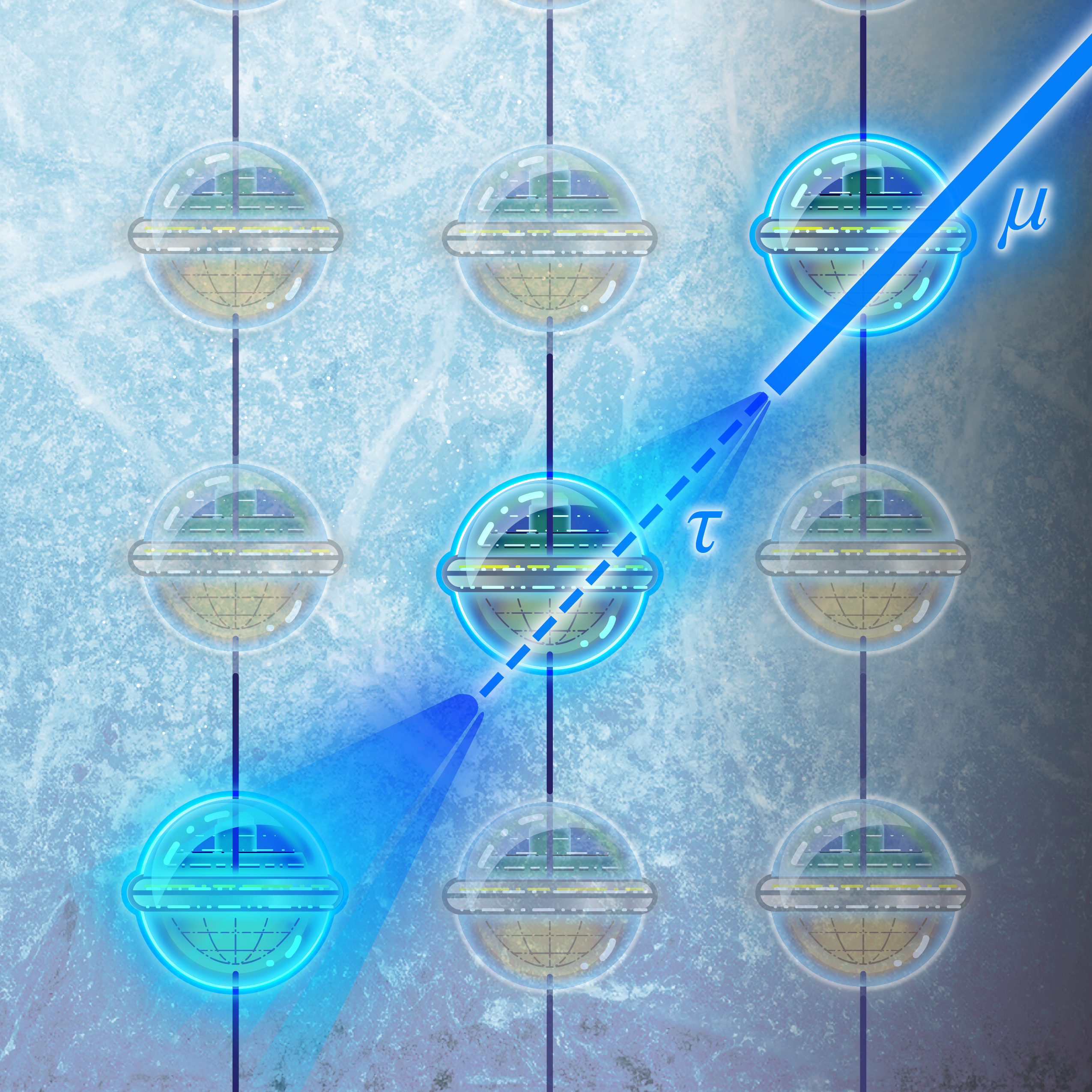}
    \captionsetup{justification=Justified}
    \caption{A schematic view of the CLFV signal topology proposed in this work --- a muon track converting to a tau track, followed by a hadronic cascade inside the detector.}
    \label{fig:Signal}
\end{figure}

\par Together with high‑energy neutrinos, interactions of cosmic rays with the Earth's atmosphere produce a large flux of energetic muons. 
At TeV energies, these muons penetrate deep into underground or under‑ice detectors and constitute the dominant background in searches for muon‑neutrino events. 
IceCube has already measured this high‑energy atmospheric muon flux~\cite{IceCube:2024pnx}. 

\par In this Letter, we propose to exploit this large sample of cosmic-muon events to search for potential signatures of CLFV at neutrino telescopes. Interestingly, the first-ever CLFV search for muon-to-electron conversion in nuclei was done using cosmic rays~\cite{Lagarrigue:1952}.    
We investigate the potential signature of muon-to-tau conversion process in IceCube detector; see Figure~\ref{fig:Signal} for a schematic view of the proposed event topology.
In neutrino telescopes, $\mu \to e$ transitions suffer from large electromagnetic shower backgrounds and poor flavor discrimination, whereas tau leptons yield more distinctive event topologies, rendering the $\mu \to \tau$ channel comparatively background-suppressed.  
Moreover, unlike $\mu \to e$, which has stringent constraints from low-energy searches~\cite{SINDRUMII:1993gxf,SINDRUMII:1998mwd,SINDRUMII:1996fti,SINDRUMII:2006dvw,MEGII:2025gzr}, $\mu \to \tau$ conversion searches~\cite{Gninenko:2001id,Sher:2003vi,Gninenko:2018num,Husek:2020fru} remain sparse and associated phenomenology far less constrained~\cite{Kanemura:2004jt, Paradisi:2005tk,Bolanos:2012zd,Abada:2016vzu,
Takeuchi:2017btl}, making this channel a promising new probe of uncharted CLFV parameter space.

\par Using existing IceCube data on cosmic‑ray–induced muons~\cite{IceCube:2024pnx}, we 
demonstrate that the existing IceCube setup already provides competitive sensitivities in the $\mu-\tau$ LFV sector and it is worth searching for this new topology. 
We further extend our analysis to future experiments like IceCube-Gen2~\cite{IceCube-Gen2:2021rkf}---the 8 km$^3$ upgraded version of IceCube, and HUNT~\cite{Chen:2026jdx}---the $30$ km$^3$ planned neutrino experiment in South China Sea, and estimate their projected sensitivities to CLFV interactions.
Our approach extends naturally to other Cherenkov neutrino telescopes such as KM3NeT~\cite{KM3Net:2016zxf}, P-ONE~\cite{P-ONE:2020ljt}, Baikal-GVD~\cite{Avrorin:2015wba},  TRIDENT~\cite{TRIDENT:2022hql}, and TAMBO~\cite{TAMBO:2025jio}. Moreover, the idea of using cosmic-ray muons, typically treated as unwanted background in neutrino experiments, as a powerful probe of new physics, can also be applied to other large-volume (near-)surface neutrino detectors, such as proto-DUNE~\cite{DUNE:2021hwx} and NOvA~\cite{NOvA:2004blv}.

\begin{figure}[t]
\centering
\includegraphics[width=0.45\textwidth]{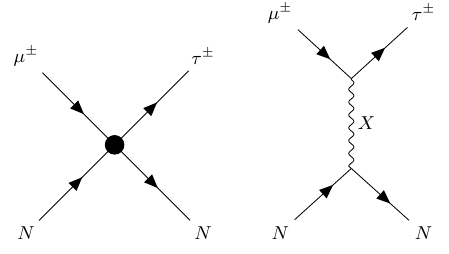}
\captionsetup{justification=Justified}
\caption{Feynman diagrams for $\mu^\pm N \rightarrow \tau^\pm N$ in the EFT approach (left) and with a mediator (right).}
\label{fig:Feyn}
\end{figure}

{\textbf {$\boldsymbol{\mu\to \tau}$ interaction channels--}}
Different channels through which $\mu \to \tau$ conversion can take place are shown in Fig.~\ref{fig:Feyn}.
The left panel shows a generic, 4-fermion EFT picture, which can be systematically described in terms of various EFT operators.
The right panel shows a model-specific realization of the EFT operator in terms of a (pseudo)scalar or (axial)vector mediator, which can be either a BSM particle, or even the SM $Z/h$ boson with LFV interactions. 

For all these scenarios, the corresponding cross-sections of $\mu^\pm N\rightarrow\tau^\pm N$ is numerically calculated using the joint pipeline of  
\texttt{Feynrules}~\cite{Alloul:2013bka}, and 
\texttt{MadGraph}~\cite{Alwall:2014hca}.
The default parton distribution functions (PDFs) of \texttt{MadGraph} are used. Some representative simulated cross-section values can be found in Supplemental Section I.\\[0.5mm]
{\textbf {Dimension-6 EFT operator.--}}  We consider an effective four-fermion operator with the general Lorentz structure~\cite{Black:2002wh} 
\begin{equation}
    \mathcal{Q}_{6\text{-dim}}=\frac{c_{\mu \tau}}{\Lambda^2}(\bar{\mu}\Gamma\tau)(\bar{q}\Gamma q),
    \label{eq:QEFT}
\end{equation}
with $\Gamma=1$, $\gamma^5$, $\gamma^\alpha$, and $\gamma^\alpha\gamma^5$  specifying scalar, pseudoscalar, vector, and axial vector interactions, respectively. Here $\Lambda$ is the relevant BSM physics scale and $c_{\mu \tau}$ the Wilson coefficients, which are set to 1 without loss of generality (by appropriate scaling of $\Lambda$). 
\\[0.5mm]
{\textbf {$Z'$-mediated interaction.--}}We also consider a concrete realization of the EFT operator in terms of an axial vector $Z'$ with LFV interactions. The Lagrangian is 
\begin{equation}
    -\mathcal{L}_{Z^\prime}\supset g_l\bar{\mu}\gamma^\alpha\gamma^5\tau Z^\prime_\alpha+g_l\bar{\nu}_\mu\gamma^\alpha\gamma^5\nu_\tau Z^\prime_\alpha+g_q \bar{q}\gamma^\alpha\gamma^5q Z^\prime_\alpha +\text{H.c.},
    \label{eq:Zp}
\end{equation}
with $g_l$ and $g_q$ being the couplings of $Z^\prime$ to leptons and quarks respectively. We have chosen these couplings to be equal in the following analysis; the CLFV sensitivities derived here only depend on the product $g_lg_q$.  The choice of axial vector $Z'$ is motivated by our EFT results, where the axial-vector-like Lorentz structure has the best sensitivity (see Fig.~\ref{fig:6dim}).
In Eq.~\eqref{eq:Zp}, we consider LFV $Z'$ couplings only to the $\mu-\tau$ sector, and flavor-conserving quark couplings. This is a phenomenological choice which reduces the parameter space while retaining the essential features relevant to our analysis. This can be justified by imposing additional symmetries in a specific ultraviolet-completion~\cite{Foot:1994vd,Altmannshofer:2016brv}. 

Our primary objective is to identify a clean and experimentally accessible signature of CLFV through $\mu \rightarrow \tau$ conversion. In this process, an on-shell tau is produced following the muon-nucleon interaction and propagates a finite distance before decaying. We consider only the cases in which the tau decays hadronically, producing a subsequent cascade in the detector. This results in a unique signal topology that can be distinguished from standard cascade-like and track-like events. In contrast, identifying $e^\pm$ signatures in IceCube is significantly more challenging due to large backgrounds and limited discrimination power. Therefore, we do not include interactions involving $Z'$ and $e^\pm$. 

Our $Z'$ analysis can be extended to other mediators. In Supplemental Section II, we discuss the results for LFV couplings of the SM $Z$ and Higgs bosons.  

{\textbf {Signal analysis.--}}
IceCube has reported the detection of \(7.92\times10^{11}\) cosmic muons over twelve years of data taking~\cite{IceCube:2024pnx}. The dataset spans a wide energy range, from $10~\mathrm{TeV}$ to more than $5~\mathrm{PeV}$, and includes events with directions $\cos\theta \ge 0.2$, where $\theta$ is the zenith angle.

The scattering of cosmic muons with ice nucleons through photon-mediated interactions leads to energy loss. Within the SM, additional weak processes involving charged-current (CC) and neutral-current (NC) interactions can also occur, although they are highly suppressed and, in all cases, preserve lepton flavor. 
In the presence of sizable CLFV, however, a muon interacting with matter may produce a final-state lepton of a different flavor, such as $\tau$ as assumed here (Fig.~\ref{fig:Feyn}). 
The number of muons converted into \(\tau\)-leptons is given by
    \begin{align}
  N_{\tau}^\text{CLFV}(E_i)=&(1-\text{BR}_{\tau \rightarrow \mu})\int_{E_\text{min}}^{E_\text{max}}dE\int_{0.2}^{1} d\text{cos}\theta\frac{d^2N_{\mu}(E)}{dE d\text{cos}\theta}\nonumber \\
  & \times \int _0^1d\eta(1-e^{-(L_{\theta}-x(E_\mu=\eta E))/\lambda(E_\mu=\eta E)})\nonumber \\
  & \times 
   \int_0^1 dy~e^{-d/l_\tau(E_\tau=y E_\mu)},
  \label{eq:wlosseff}
\end{align}
where $\frac{d^{2}N_{\mu}}{dE\, d\cos\theta}$ denotes the muon flux reaching the detector, inferred from IceCube measurements~\cite{IceCube:2024pnx} after all selection cuts are applied, expressed as a function of the true energy $E$. The correlation between the true energy $E$ and the reconstructed energy $E_{i}$ is provided in Ref.~\cite{IceCube:2024pnx}. We integrate this flux over the full true-energy range. For the zenith angle, we only consider $0.2<\cos\theta <1$, as reported by IceCube.  
The flux is weighted by the probability that a muon undergoes the CLFV transition, which depends on the distance $L_{\theta}$ that the muon travels through the ice with $\theta=0\degree$ corresponding to the muons traveling through the height of the detector. The mean free path for the CLFV process, $\lambda(E_\mu) = 1/(n\,\sigma(E_\mu))$, is determined by the muon cross section $\sigma(E_{\mu})$ and the number density of nucleons in ice, $n \approx 1~\mathrm{g\,cm^{-3}}/m_{\text{nucleon}}$.

Muons lose energy as they propagate through matter via four main processes: ionization, bremsstrahlung, pair production, and photonuclear interactions. The integration over the variable $\eta$ in Eq.~\eqref{eq:wlosseff} incorporates this energy loss during propagation. Empirically, the mean  energy loss over a column-distance $x$ can be described by the relation
\begin{equation}
    \frac{dE_\mu}{dx} = -a - bE_\mu,
    \label{loss-empiric}
\end{equation}
where $a$ represents the ionization energy loss and $b$ accounts for all radiative processes. Although both $a$ and $b$ are, in principle, energy dependent, they can be treated as constants at high energies, with 
$a = 2.4~\mathrm{MeV~g^{-1}~cm^{2}}$ and  $b = 3.2\times 10^{-5}~\mathrm{g^{-1}~cm^{2}}$~\cite{IceCube:2012iea}. For muon energies above $E_\mu \gtrsim 75~\mathrm{GeV}$, radiative processes dominate the overall energy loss. 

The taus are reconstructed in the detector as cascade-like events. This topology is the same as that produced by CC electron-neutrino interactions and by NC neutrino interactions. To distinguish tau-induced cascades from these background events, we require the tau to travel at least some distance, $d$, before its decay (see Eq.~\eqref{eq:wlosseff}). We therefore integrate over the muon inelasticity, $y$,
defined as the fraction of the muon's energy transferred to the tau, weighted by the probability that the tau travels farther than $d$, given by $\exp\!\left[-d/l_\tau(E_\tau = y\,E_\mu)\right]$
where $l_\tau$ is the decay length of the tau in the lab frame. In this work, we take $d=3$\,m, motivated by the maximum mean distance traveled by $D$ mesons produced in hadronic interactions~\cite{jin_2024_13235037}.  
The number of cascades generated by taus is further suppressed by the branching ratio for taus decaying into muons, $\text{BR}_{\tau \rightarrow \mu}$.

{\textbf {Background analysis.--}} At very high energies, muon stochastic energy losses can produce signatures in the detector that resemble cascade-like events. These constitute the main background to the tau search, as the detector is dominated by the flux of cosmic muons. The number of cascade events induced by muon energy losses is given by
\begin{align}
  N_{\tau}^\text{bkg}(E_i,\eta_\text{req})&=\int_{E_\text{min}}^{E_\text{max}}dE\int_{0.2}^{1} d\text{cos}\theta\frac{d^2N_{\mu}(E)}{dE d\text{cos}\theta}P(E,\eta_\text{req}),
  \label{eq:bkg}
\end{align}
where $P(E,\eta_{\text{req}})$ represents the probability that a muon loses most of its energy in its first interaction inside the detector, leaving a residual muon with energy below $100~\mathrm{GeV}$. To estimate the muon energy deposition along its path through the detector, we simulate muon propagation using \texttt{PROPOSAL}~\cite{Koehne:2013gpa}; see Supplemental  Section III for more details.

{\textbf {Results.--}} To constrain the new physics, we compute the following test-statistics, $\mathcal{TS}$, based on Poisson log-likelihood:
\begin{figure*}[t]
\centering
    \includegraphics[width=\linewidth]{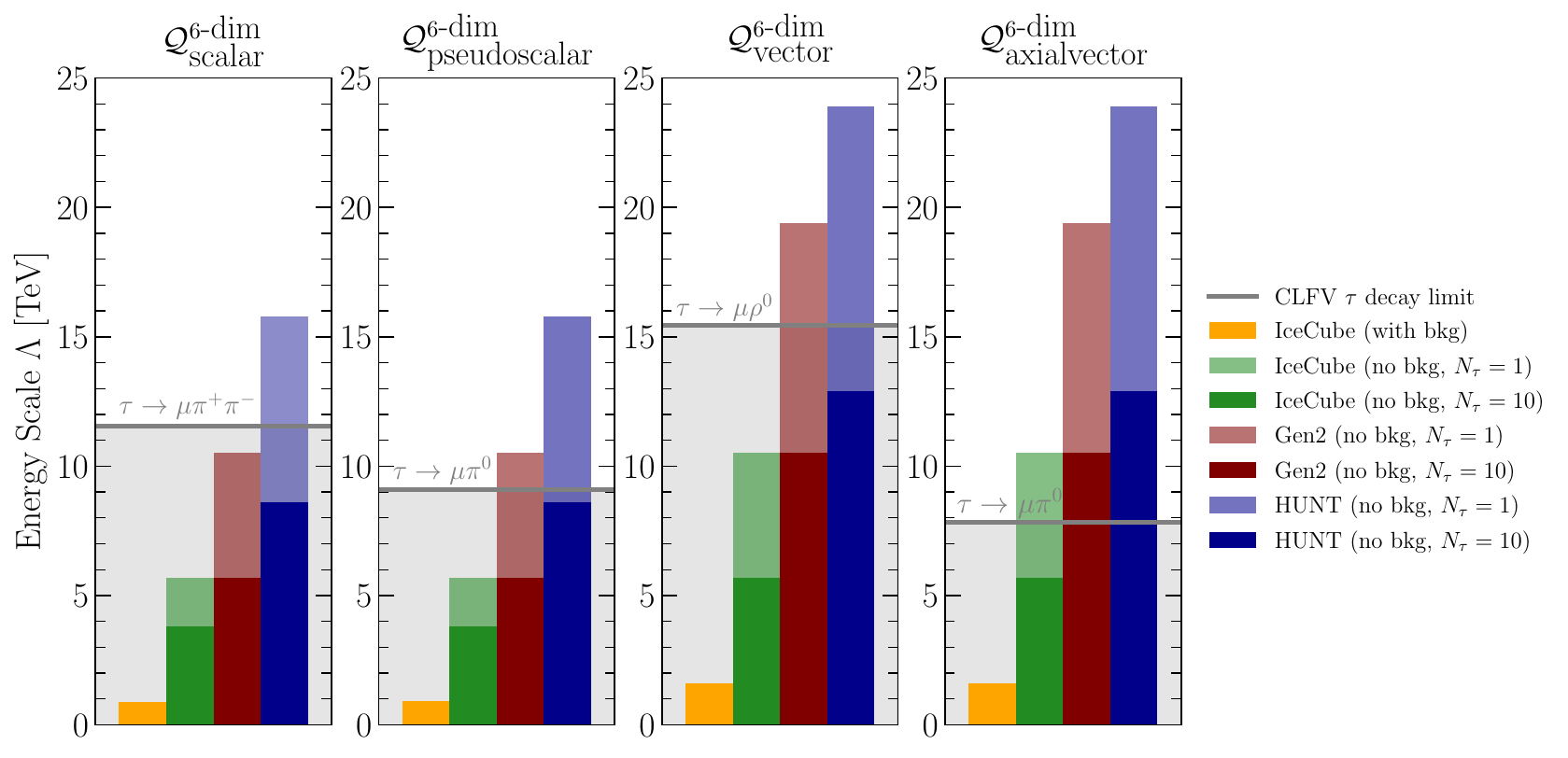}
    \captionsetup{justification=Justified}
    \caption{The 95\% CL sensitivity on the new physics scale, $\Lambda$, obtained for various EFT operators; see text for details. 
    The gray lines show the current $\tau$ LFV constraints.}
    \label{fig:6dim}
\end{figure*}
\begin{align}
    \mathcal{TS}=& 2\sum_i \left(N_{\tau}^\text{CLFV}(E_i)+N_{\tau}^\text{bkg}(E_i)\right)-N_{\tau}^\text{bkg}(E_i)\nonumber \\
    & +N_{\tau}^\text{bkg}(E_i)\text{log}\left(\frac{N_{\tau}^\text{bkg}(E_i)}{N_{\tau}^\text{CLFV}(E_i)+N_{\tau}^\text{bkg}(E_i)}\right),
\end{align}
where we assume $\mathcal{TS}$ follows a $\chi^2$-distribution with one degree of freedom given by the CLFV parameter strength.
Requiring that the CLFV-induced $\tau$ events be distinguishable from background at 95\% confidence level (CL) yields the sensitivities presented in the following subsections.
In what follows, when we impose sensitivities, we assume these CLFV events are absent from the IceCube data, though we note that no dedicated search has been performed in the experiment looking for this event topology.

We have also obtained sensitivities for next-generation neutrino experiments like IceCube-Gen2 and HUNT.
To obtain these, Eq.~\eqref{eq:wlosseff} is used with an additional multiplicative factor, corresponding to the excess numbers of muons detected by these larger-volume detectors.
For simplicity, we have chosen this multiplicative factor to be the ratio of their detector volume to IceCube's. 
Additionally, the dimensions corresponding to these experiments are adjusted in Eq.~\eqref{eq:wlosseff}.
The IceCube-Gen2 and HUNT projections are obtained under a no-background assumption by requiring at least one or ten CLFV-induced $\tau$ events.
The scales reached under this assumption correspond to the sensitivity at which one or ten CLFV-induced $\tau$ events would be expected; they should be interpreted as the reach attainable with sufficiently improved background rejection, not as constraints.
For comparison, the same no-background sensitivity is also evaluated for IceCube using existing data. \\[0.5mm]
{\textbf {EFT case.--}} Following the prescription described, sensitivities on the new physics scale of the different EFT operators~\eqref{eq:QEFT} are evaluated and depicted in Fig.~\ref{fig:6dim}. 
The orange bin exhibits the $95\%$ CL sensitivity on $\Lambda$ obtained from the existing IceCube data with simulated background analysis. 
The green-colored bin shows the sensitivity reach in $\Lambda$ from the same IceCube data under a no-background assumption (i.e., the scale below which at least one or ten CLFV-induced $\tau$ events would be expected, for the lighter and darker green respectively). It represents the ultimate reach of this analysis if background-rejection methods sufficient to suppress the muon-induced cascade background can be developed, rather than a current constraint.
The light shaded green bin and the darker shaded green bin correspond to the requirement of one and ten CLFV-induced $\tau$ events, respectively.
Requiring a larger number of CLFV-induced $\tau$ events naturally weakens the sensitivity reach in $\Lambda$.
The same no-background sensitivity analysis is also performed for IceCube-Gen2 and HUNT; the resulting projections are shown in maroon and dark blue, respectively.

The leptonic and semi-leptonic LFV tau decays~\cite{ParticleDataGroup:2024cfk} set stringent constraints on the new physics energy scale for the EFT operators considered here~\cite{Black:2002wh}.
The specific decay channel that gives the most stringent constraint depends on the Lorentz structure used in the EFT operator~\eqref{eq:QEFT}. 
The various LFV tau-decay branching ratios (BRs) used in this analysis are tabulated in Table~\ref{tab:BRLFV} in Supplemental Section IV, along with a brief overview of obtaining those constraints.

\begin{figure}[t]
    \centering
    \includegraphics[scale=0.45]{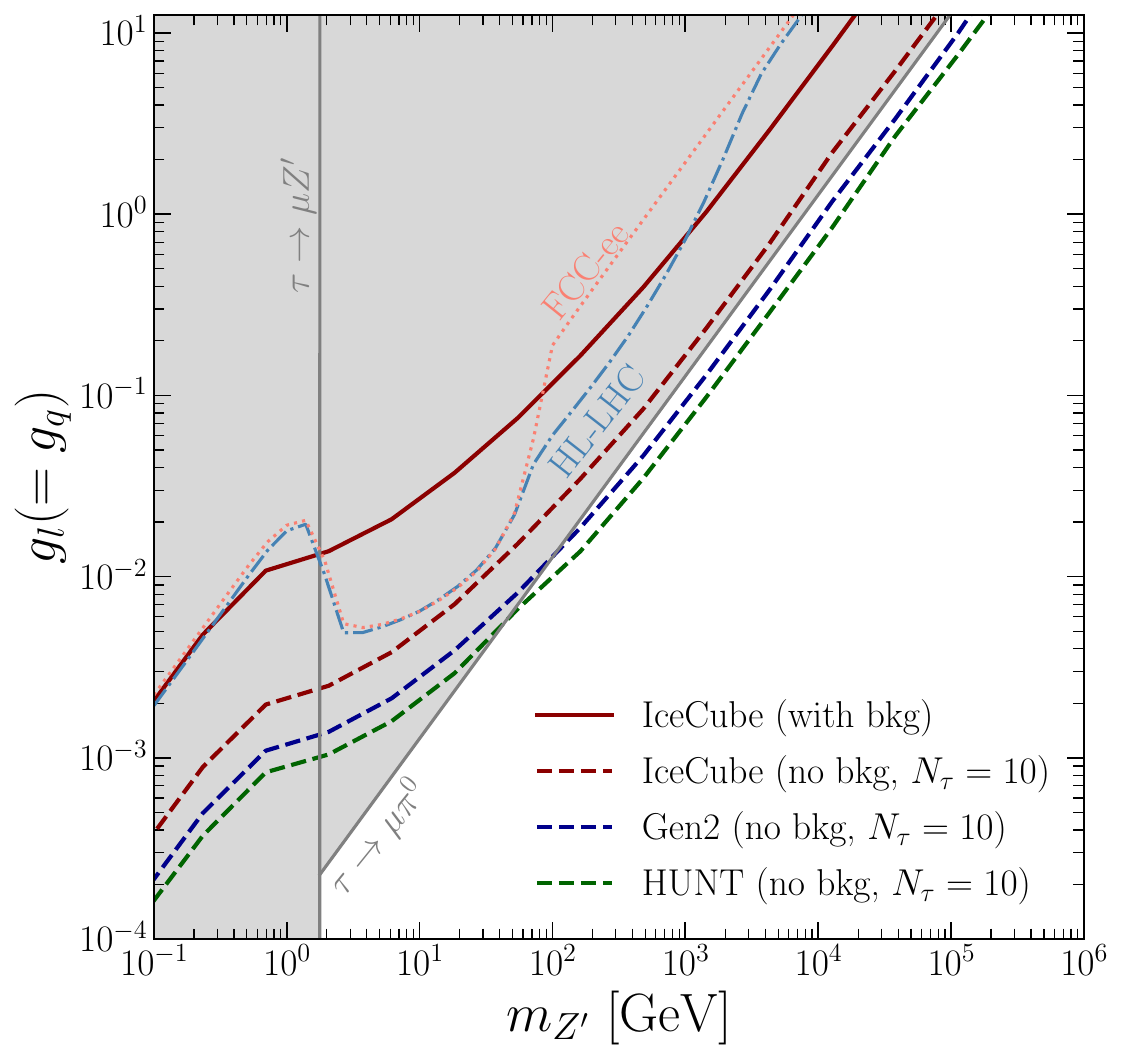}
    \captionsetup{justification=Justified}
    \caption{Sensitivity reach on the CLFV parameter space for an axial vector $Z'$; see text for details. The shaded regions are excluded by $\tau$ LFV searches. 
    }
    \label{fig:Zp}
\end{figure}

The upper limit on BR($\tau\to \mu\pi^0$) is one-order magnitude weaker than BR($\tau \to\mu\rho^0$) and BR($\tau\to\mu\pi^\pm$) (see Table~\ref{tab:BRLFV}).
On the other hand, the CLFV cross-section for scalar and pseudoscalar operators is $\mathcal{O}(10)$ times lower compared to that for vector and axial vector operators for the same energy scale, $\Lambda$.
The combined effect of the difference in LFV $\tau$ decay BR upper limits and the variation in CLFV cross-section for different EFT operators for the same $\Lambda$ makes the axial-vector case most promising for our signal (Fig.~\ref{fig:6dim}). \\[0.5mm]
{\textbf {$Z^\prime$ case.--}} 
For the $Z'$ model given by Eq.~\eqref{eq:Zp}, we derive the IceCube sensitivity in the $Z'$ mass-coupling parameter space in Fig.~\ref{fig:Zp} with the current data and including our background treatment (dark red solid curve).
The no-background analysis of IceCube data is sensitive to even larger scales (dark red dot-dashed curve).
The projected sensitivities for next-generation experiments IceCube-Gen2 and HUNT (also obtained under a no-background assumption) are shown as blue and green dashed lines, respectively.

The two solid gray lines in Fig.~\ref{fig:Zp} denote the two strongest lab constraints coming from $\tau\to \mu\pi^0$ search at BaBar~\cite{BaBar:2006jhm} and $\tau\to \mu+{\rm invisible}$ at Belle~\cite{Belle:2025bpu}.
The latter constraint is only applicable only when $m_{Z'}<m_\tau-m_\mu$ (the vertical line in Fig.~\ref{fig:Zp}). There are various other constraints on the parameter space shown in Fig.~\ref{fig:Zp}; however, they are weaker than those discussed above, and therefore, are not shown here. See Supplemental  Section V for details.

For comparison, the salmon dotted and steel-blue dot-dashed curves, on the other hand, are the projected sensitivities from future collider experiments FCC-ee~\cite{FCC:2018evy} and high-luminosity LHC (HL-LHC)~\cite{CidVidal:2018eel}, respectively. 
These lines are obtained from a simple estimate of $3\sigma$ sensitivity of same-sign di-lepton pair production  ($\mu^\pm\mu^\pm\tau^\pm\tau^\pm$), following Ref.~\cite{Altmannshofer:2016brv}. It is remarkable that even the existing IceCube setup gives us a sensitivity comparable to those from future colliders. While the IceCube sensitivity still remains within the currently excluded region, Gen-2 and HUNT can probe new parameter space for $m_{Z'}\gtrsim100~\mathrm{GeV}$.

{\textbf {Conclusions.--}}
Cosmic muons are abundantly detected by neutrino telescopes and constitute the main background in neutrino identification.
In this Letter, we have shown that they can be exploited as a high-energy probe of CLFV through $\mu \rightarrow \tau$ conversion.
Assuming that no event of this nature is present in the IceCube data, we establish the sensitive range of IceCube to these CLFV  operators.

These results rely on background discrimination. 
Additional information encoded in the event topology, which is not considered in this work, can be further exploited to enhance rejection performance. 
In particular, the nanosecond-level timing resolution of IceCube DOMs~\cite{IceCube:2016zyt} provides an additional handle for improving event identification. 
Beyond this, machine learning techniques offer a promising and robust framework for fully leveraging the rich topological and temporal information of events~\cite{Wille:2019pub,IceCube:2024nhk}.
Should such techniques succeed in suppressing the dominant muon-induced cascade background, IceCube's reach with existing data could approach the no-background sensitivity results shown in Figs.~\ref{fig:6dim} and~\ref{fig:Zp}.

To conclude, our results highlight a strong complementarity between neutrino telescopes and collider experiments, particularly at a time when the choice of next generation collider is still under discussion. 
Finally, we encourage ongoing experiments such as IceCube and KM3NeT to look for these signal topologies in existing data. 

\textbf {Acknowledgments--}
We thank Max Fieg and Sebastian Trojanowski for useful discussions. The work of WM and BD was partly supported by the U.S. Department of Energy under grant No.~DE-SC0017987. BD was also partly supported by a Humboldt Fellowship from the Alexander von Humboldt Foundation. 
CAA are supported by the Faculty of Arts and Sciences of Harvard University, the National Science Foundation, the Canadian Institute for Advanced Research, the Research Corporation for Science Advancement, the John Templeton Foundation, and the David \& Lucile Packard Foundation. IMS is supported by STFC grant ST/T001011/1. 
MS acknowledges support from the Early Career Research Grant by Anusandhan National Research Foundation (project number ANRF/ECRG/2024/000522/PMS). BD thanks the organizers of CLFV 2026 at Sapienza Universit\'{a} di Roma for local hospitality during the final stages of the work.

\bibliography{ref}
\clearpage



\newpage

\onecolumngrid
\appendix

\setcounter{equation}{0}
\setcounter{figure}{0}
\setcounter{table}{0}
\setcounter{section}{0}
\renewcommand{\theequation}{S\arabic{equation}}
\renewcommand{\thefigure}{S\arabic{figure}}
\renewcommand{\thetable}{S\arabic{table}}
\renewcommand{\thesection}{S\arabic{section}}

\section{\Large{Supplemental Material}}

\section{I. $\mu^\pm N \rightarrow \tau^\pm N$ cross-section}
\begin{figure}[h!]
    \centering
    \includegraphics[height=7cm,width=10cm]{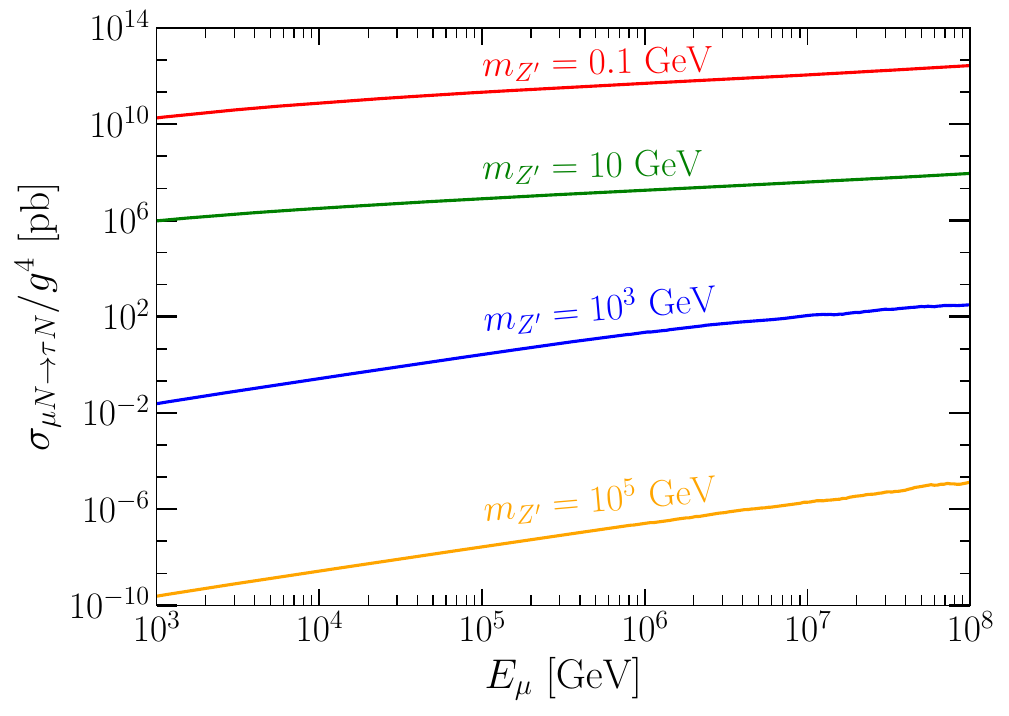}
    \captionsetup{justification=Justified}
\caption{$\mu^\pm N \rightarrow \tau^\pm N$ cross-section, normalized by the coupling of $Z^\prime$ to $\mu-\tau$ and quarks ($g_l=g_q\equiv g$ in Eq.~\eqref{eq:Zp}) as a function of incoming muon energy for different $Z^\prime$ masses.}
\label{Mdg-CS}
\end{figure}

The neutral current (NC) deep inelastic cross-section of $\mu^\pm N \rightarrow \tau^\pm N$ (where $N=p,n$ is the nucleon, with $p$ and $n$ being protons and neutrons) mediated by $Z^\prime$, and normalized by the $Z'$ gauge coupling $g_l=g_q\equiv g$, is calculated numerically using \texttt{MadGraph}~\cite{Alwall:2014hca} and is shown in Fig.~\ref{Mdg-CS} as a function of the incoming muon energy for different $Z'$ masses. 
\section{II. SM $Z$ and Higgs-mediated CLFV interactions}
\label{App:ZH}
\begin{figure}[h!]
\centering
    \includegraphics[scale=0.6]{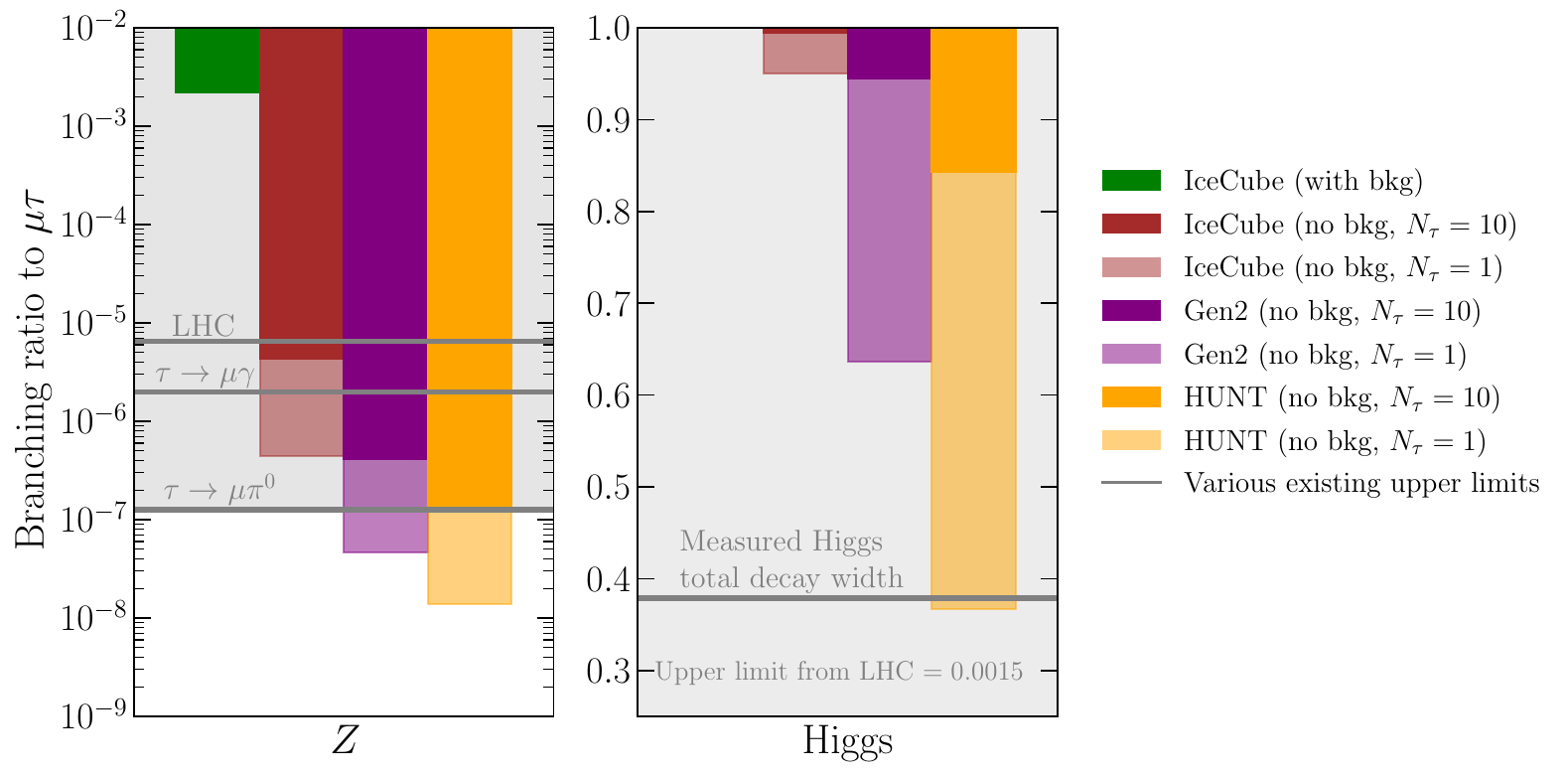}
    \captionsetup{justification=Justified}
    \caption{Constraints and sensitivities on the branching ratio of $Z$ and Higgs decaying into $\mu$ and $\tau$. The green band is the constraint from the background-informed analysis of existing IceCube data; the red, purple, and orange bands show the sensitivity reach (no-background assumption) for IceCube, IceCube-Gen2, and HUNT, respectively. See Appendix~\ref{App:ZH} for details. }
    \label{fig:BRZH}
\end{figure}
If SM $Z$ or Higgs boson has an effective CLFV interaction in the $\mu-\tau$ sector (coming from some BSM physics), it can also mediate the $\mu\to \tau$ transition in Fig.~\ref{fig:Feyn}.  
The analysis here is very similar to the $Z'$ analysis done in the main text, but with the $Z'$ quark couplings replaced by the SM $Z$ and Higgs coupling to quarks, while on the leptonic side, we assume the $Z'\mu\tau$ and $h\mu\tau$ LFV couplings, and find the IceCube sensitivity to this coupling, which can be translated into an upper limit on the corresponding decay BRs, as shown in Fig.~\ref{fig:BRZH}. The prescription of obtaining these sensitivities is same as mentioned in the main text. The green band in Fig.~\ref{fig:BRZH} corresponds to the upper limit on BR$_{\mu\tau}$ attained from the $\chi^2$ analysis of the existing IceCube data with background signal being taken into account. The red, purple, and orange bands correspond to the sensitivity reach in BR$_{\mu\tau}$ from IceCube, IceCube-Gen2, and HUNT, respectively, under a no-background assumption; these indicate the reach achievable if background-rejection methods sufficient to suppress the muon-induced cascade background can be developed. The darker and lighter bins in this context denote the requirement of minimum number of CLFV-induced tau events to be ten and one,  respectively. 

The CLFV interaction is mediated here by the axial vector part of $Z$, and hence, the strongest lab constraint comes from the non-observation of $\tau \rightarrow \mu\pi^0$~\cite{BaBar:2006jhm}. There is also an LFV constraint from $\tau \to \mu\gamma$, because we have diagonal $Z$ and Higgs couplings to $\mu$ and $\tau$ in the SM. This is unlike in the EFT and $Z'$ cases shown in Figs.~\ref{fig:6dim} and~\ref{fig:Zp}, where we only assumed the LFV $\mu\tau$ coupling in the lepton sector, and therefore, the $\tau\to \mu\gamma$ constraint did not apply. 

In addition, we have direct LHC constraints. Both ATLAS and CMS experiments at the LHC have searched for $Z\to \mu\tau$~\cite{ATLAS:2021bdj, CMS:2025wqy} as well as $h\to \mu\tau$ decays~\cite{CMS:2021rsq, ATLAS:2023mvd}. As shown in Fig.~\ref{fig:BRZH} left panel, the LHC constraint is weaker than the low-energy tau LFV constraints, and is comparable to our IceCube sensitivity with no background, while the Gen2 and HUNT sensitivities can surpass the current constraints.

However, in the Higgs case (Fig.~\ref{fig:BRZH} right panel), the IceCube sensitivities are not competitive. This is primarily due to the suppressed Higgs Yukawa coupling to first-generation quarks, which greatly suppresses the $\mu N\to \tau N$ cross-section mediated by the Higgs. In this case, even the LHC constraint on BR($h\to \mu\tau$) coming from the difference between the  measured total Higgs decay width~\cite{CMS:2026igg} and the SM prediction~\cite{LHCHiggsCrossSectionWorkingGroup:2016ypw} gives a stronger constraint than our proposed CLFV channel. 
\section{III. Details about background event simulation}
\label{App:Proposal}
\begin{figure}[t]
    \centering
    \includegraphics[width=0.6\linewidth]{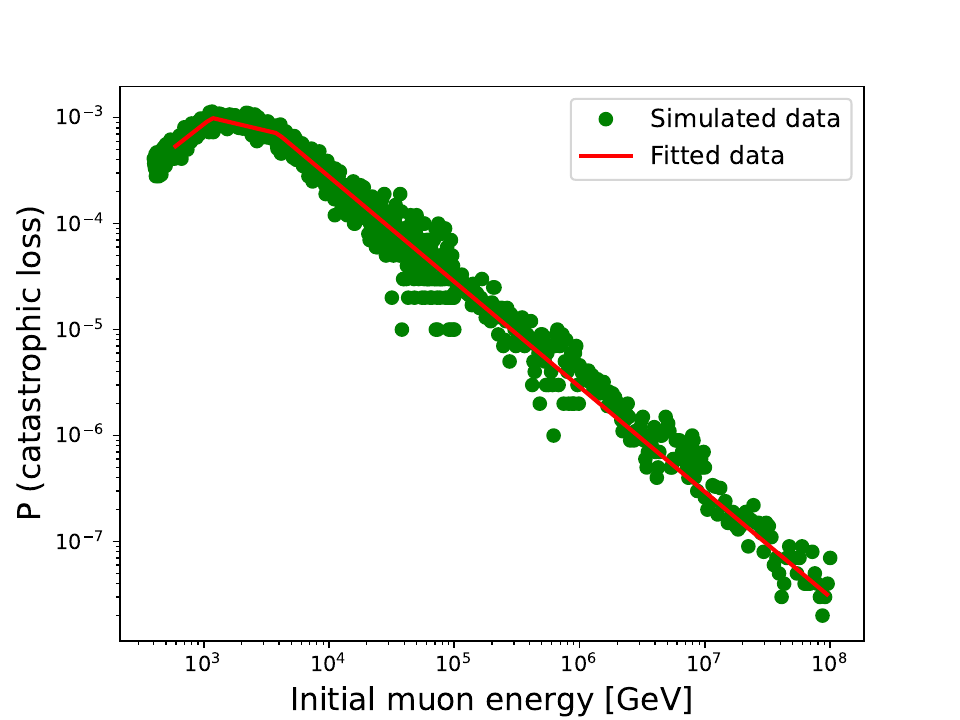}
    \captionsetup{justification=Justified}
    \caption{Probabilities of muons undergoing stochastic energy loss as a function of the initial muon energy. The data points are simulated using the \texttt{PROPOSAL} code. The fitted function is given in Eq.~\eqref{eq:fitted}, which is used to estimate the muon-induced cascade background in our analysis.}
    \label{fig:catloss}
\end{figure}
We used {\tt PROPOSAL}~\cite{Koehne:2013gpa} to simulate $10^8$ events for each initial muon energy considered in our analysis and calculated the probability of muons undergoing stochastic energy loss, $P_\text{cat-loss}(E,\eta_\text{req})$, as they propagate in the IceCube detector. A muon is considered to lose energy stochastically when a required fraction of its initial energy, $\eta_\text{req}$, is lost during its propagation. A threshold of $100$ GeV is also implemented in this simulation to ensure that the muon after its stochastic energy loss is not capable of lighting any PMT. The obtained probability of muons' stochastic energy loss is then smoothed by fitting the data points with a reasonable function (Fig.~\ref{fig:catloss}). The fitted function is a broken power law function that looks like:
\begin{align}
P_\text{cat-loss}(E,\eta_\text{req})=\left\{
\begin{array}{ll}
       N_1E^{k_1} & (E\le E_{\text{break},1}),\\
       N_2E^{k_2} &(E_{\text{break},1}<E\le E_{\text{break},2}),\\
      N_3E^{k_3} &(E> E_{\text{break},2}).
    \end{array}\right. 
    \label{eq:fitted}
\end{align}
For our work, we have chosen $\eta_\text{req}$ to be $0.9$. In other words, we call a muon's energy loss catastrophic when it loses $90\%$ of its initial energy. In principle, $P_\text{cat-loss}$ is also a function of the distance that muon travels inside the detector. To make the simulation computationally more effective, we have taken the angle-averaged distance of all the possible distances muons can travel inside IceCube. The fitting parameters for our chosen scenario have turned out to be: $N_1=1.5\times 10^{-6}$, $k_1=0.9$, $N_2=0.007$, $k_2=-0.28$, $N_3=2.6$ and $k_3=-0.99$.

\begin{figure}
    \centering
    \includegraphics[width=0.6\linewidth]{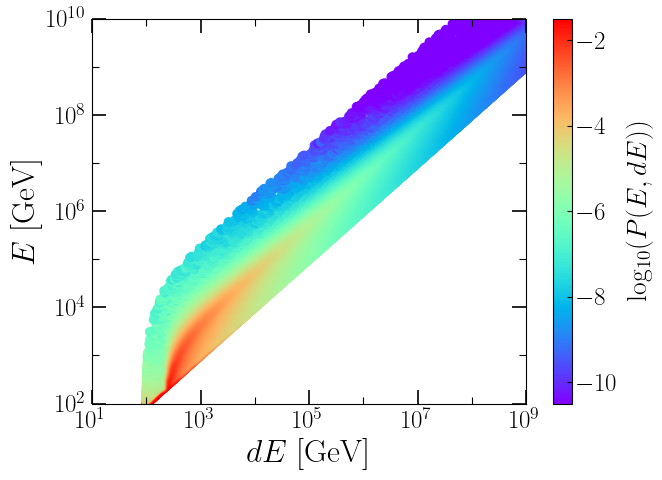}
    \captionsetup{justification=Justified}
    \caption{The heatmap of probability of muons losing energy catastrophically while crossing $1$ km of ice inside the IceCube detector as a function of the initial energy of these muons and the energy it loses during this propagation.}
    \label{fig:PdE}
\end{figure}

The probability of muons losing energy stochastically while passing through $1$ km of IceCube detector as a function of the muon energy and the amount of energy it loses for a threshold energy of $100$ GeV is shown in Fig.~\ref{fig:PdE} which matches quite well with the ones in Refs.~\cite{Robertson:2018,Hill:2020kmk}. These probabilities are again calculated using the \texttt{PROPOSAL} code. A muon with initial energy $E_\mu$ cannot have an energy loss more than $E_\mu$ and that sets the slope of the heatmap on its right side.
\section{IV. Decay width of various LFV $\tau$-decay channels}
\label{App:Opconstraint}
\begin{figure}[t]
\centering
\includegraphics[width=0.99\textwidth]{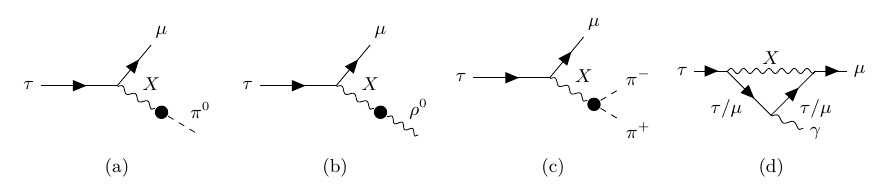}
\captionsetup{justification=Justified}
\caption{Feynman diagrams for  various  tau LFV decay channels: (a) $\tau\rightarrow\mu\pi^0$, (b) $\tau\rightarrow\mu\rho^0$, (c) $\tau\rightarrow\mu\pi^+\pi^-$ and (d) $\tau\rightarrow\mu\gamma$, all mediated by $X$ that has a $\mu\tau$ LFV coupling and coupling to quarks (as in Fig.~\ref{fig:Feyn} right panel).}
\label{fig:FeynCLFV}
\end{figure}

The same $\mu\tau$ effective coupling that leads to the $\mu\to\tau$ conversion signal discussed in main text will also lead to various low-energy tau LFV processes, which we briefly discuss in this section. The Feynman diagrams of these processes are shown in Fig.~\ref{fig:FeynCLFV}. The decay widths of such decay processes depend on the Lorentz structure of the concerned EFT operators or the nature of the mediator. 

If the EFT operator has an axial-vector type Lorentz structure, the strongest constraint on $\Lambda$ comes from the upper limit on ${\rm BR}(\tau\to \mu\pi^0)<1.1\times 10^{-7}$ from BaBar~\cite{BaBar:2006jhm}. The decay width of this process with the LFV vertex taking the $\gamma^\mu\gamma^5$ Lorentz structure takes the following form~\cite{Black:2002wh}:
\begin{equation}
\Gamma \left( \tau \rightarrow \mu \pi^0 \right)_{\rm axialvector} = 
\frac{1}{32\pi\Lambda^4} f_\pi^2 m_\tau^3\left(1-\dfrac{m_\pi^2}{m_\tau^2}\right)^2  \,,
\label{Eq:OpAxial}
\end{equation} 
with $m_\tau$ and $m_\pi$ being the $\tau$ and pion mass respectively and $f_\pi$ is the pion decay constant whose value is taken to be $130.2$ MeV~\cite{ParticleDataGroup:2024cfk}.

The same $\tau\rightarrow\mu\pi^0$ channel also gives the most stringent bound on the new physics scale if the concerned EFT operator has pseudoscalar type Lorentz structure. The decay width of $\tau \rightarrow \mu \pi^0 $ under the $\gamma^5$ Lorentz structure for the LFV vertex goes like~\cite{Black:2002wh}:
\begin{equation}
\Gamma (\tau \rightarrow \mu \pi^0)_{\rm pseudoscalar} 
=  \dfrac{1}{32\pi\Lambda^4}\dfrac{ f_\pi^2 m_\pi^4}{(m_u+m_d)^2} m_\tau\left(1-\dfrac{m_\pi^2}{m_\tau^2}\right)^2,
\end{equation}
with $m_u$ and $m_d$ being the up and down quark masses, respectively.

Now, if the EFT operator has a vector type Lorentz structure, the strongest constraint on $\Lambda$ comes from ${\rm BR}(\tau\to \mu\rho^0)<1.7\times 10^{-8}$ from Belle~\cite{Belle:2023ziz}. The decay width of this process with the LFV vertex having $\gamma^\mu$ Lorentz structure has the following form~\cite{Black:2002wh}:
\begin{equation}
\Gamma(\tau \rightarrow \mu \rho^0)_{\rm vector} = 
\frac{ 1}{32\pi \Lambda^4} f_\rho^2 m_\tau^3{\left( 1 -
\frac{m_\rho^2}{m_\tau^2} \right)}^2 {\left( 1 + 2  \frac{m_\rho^2}{m_\tau^2} 
\right)}  ,
\end{equation} 
with $m_\rho$ being the mass of $\rho^0$ meson and $f_\rho$ being the $\rho$ decay constant whose value is chosen to be  $208.5$ MeV~\cite{Sun:2015enu}.

On the other hand, if the relevant EFT operator has scalar type Lorentz structure, then the most stringent bound on the energy scale $\Lambda$ comes from ${\rm BR}(\tau \rightarrow \mu \pi^+ \pi^-)<2.1\times 10^{-8}$ from Belle~\cite{Belle:2012unr}. The differential decay width of this process looks like~\cite{Black:2002wh}:
\begin{equation}
    d\Gamma(\tau \rightarrow \mu \pi^+ \pi^-)_{\rm scalar}= 2d\Gamma(\tau \rightarrow \mu \pi^0 \pi^0)\equiv\dfrac{1}{64\pi^3\Lambda^4}\dfrac{m_\pi^4}{(m_u+m_d)^2}E_1dE_1dE_2,
\end{equation}
with $E_1$ and $E_2$ being the energies of tau and muon,  respectively. 

Table~\ref{tab:BRLFV} summarizes the most relevant tau LFV process for each Lorentz structure type, the corresponding experimental upper limit on its BR, and the lower limit on the new physics scale $\Lambda$. 
\begin{center}
\begin{table}[h!]
    \begin{tabular}{|p{4.6cm}|c|c|c|c|}
    \hline
         & $1$ & $\gamma^5$ & $\gamma^\alpha$ & $\gamma^\alpha\gamma^5$  \\
         \hline
        Dominant CLFV process  & $\tau\rightarrow\mu\pi^+\pi^-$ & $\tau\rightarrow\mu\pi^0$ & $\tau\rightarrow\mu\rho$ & $\tau\rightarrow\mu\pi^0$ \\
        \hline
        Upper Limit on Branching Ratio & $2.1\times 10^{-8}$~\cite{Belle:2012unr} & $1.1\times 10^{-7}$~\cite{BaBar:2006jhm} & $1.7\times 10^{-8}$~\cite{Belle:2023ziz} & $1.1\times 10^{-7}$~\cite{BaBar:2006jhm}\\
        \hline
        Lower limit on $\Lambda$ & $11.6$ TeV & $9.6$ TeV & $15.4$ TeV & $7.9$ TeV \\
        \hline
    \end{tabular}
    \captionsetup{justification=Justified}
    \caption{Upper limits on the BRs of different $\tau$ LFV decays used in this work~\cite{ParticleDataGroup:2024cfk} and the lower limits on $\Lambda$ obtained from those values for the different relevant Lorentz structures of the dimension-6 EFT operator in Eq.~\eqref{eq:QEFT}. These bounds are updated from Ref.~\cite{Black:2002wh} taking into account the improved experimental results for the BR limits.}
    \label{tab:BRLFV}
\end{table}
\end{center}

Additionally, there is another $\tau$ LFV decay channel where tau can decay to muon and photon. For this particular decay channel, the EFT operator or mediator has to simultaneously have the off-diagonal $\mu\tau$ coupling and diagonal $\mu\mu$ or $\tau\tau$ coupling. 
For the scenarios considered here, this is only applicable to the $Z$ boson case with an additional LFV $\mu\tau$ coupling. The BR of $\tau$ decaying to muon plus photon under this assumption takes the following form~\cite{Lindner:2016bgg}:
\begin{equation}
    \text{BR}({\tau\rightarrow\mu\gamma})=\dfrac{3(4\pi)^3\alpha_\text{EM}}{4G_F^2}\left(|A_\text{M}^{\tau\mu}|^2+|A_\text{E}^{\tau\mu}|^2\right),
\end{equation}
with $\alpha_\text{EM}$ and $G_F$ being the fine structure constant and Fermi constant respectively. If $g_{V(A)}^{l_il_j}$ corresponds to the coupling of $i$-th flavor lepton ($l_i$) and $j$-th flavor lepton ($l_j$) where the underlying Lorentz structure is vector (axial vector), 
\begin{align}
\label{eq:AEMtaumugamma}
   |A_\text{M}^{\tau\mu}|=& -\dfrac{1}{(4\pi)^2}\sum_{l=\mu,\tau} g_V^{l\mu*}g_V^{l\tau}I_{l}^{++}+g_A^{l\mu*}g_A^{l\tau}I_{l}^{+-},  \nonumber \\
 |A_\text{E}^{\tau\mu}|=&\dfrac{i}{(4\pi)^2}\sum_{l=\mu,\tau} g_A^{l\mu*}g_V^{l\tau}I_{l}^{-+}+g_V^{l\mu*}g_A^{l\tau}I_{l}^{--},
\end{align}
and the integrals $I_{l}^{\pm\pm}$ for a specific lepton $l$ can be found in Appendix A of Ref.~\cite{Lindner:2016bgg}. In the approximation of mediator mass ($Z$ boson mass, $m_Z$, in this case) being much larger compared to the lepton mass ($m_l$), it can be written as:
\begin{equation}
    I_l^{(\pm)_1,(\pm)_2} \simeq \dfrac{1}{m_Z^2} \dfrac{2}{3}\left[1+(\pm)_1 \dfrac{m_\mu}{m_\tau} - 3(\pm)_2 \dfrac{m_l}{m_\tau}\right].
\end{equation}
For our model, $l$ can be only $\mu$ and $\tau$, $g^{\mu\tau}_V=0$ and $g^{\mu\tau}_A=g$ where $g$ is the coupling of $Z$ with muon and tau. $g^{\mu\mu}_{V(A)}$ and $g^{\tau\tau}_{V(A)}$ in Eq. \eqref{eq:AEMtaumugamma} are the SM vector (axial vector) coupling of $Z$ boson with the leptons, $g^\text{SM}_{Z,V(A)}$\footnote{\parbox[t]{0.98\textwidth}{The SM Lagrangian of $Z$ boson interacting with muon and tau can be written as $\mathcal{L}^\text{SM}_Z \supset (g^\text{SM}_{Z,V}\gamma^\alpha - g^\text{SM}_{Z,A}\gamma^\alpha \gamma^5) Z_\alpha(\mu\bar{\mu}+\tau\bar{\tau})$.  Here $g^\text{SM}_{Z,V}=g_Z(-1/4+\sin^2\theta_w)$ and $g^\text{SM}_{Z,A}=-g_Z/4$ with $\theta_w$ being the weak mixing angle and $g_Z=g_2/\cos\theta_w$ ($g_2$ is the $SU(2)_{L}$ gauge coupling in the SM).}}. This further simplifies the Eq. \eqref{eq:AEMtaumugamma} into
\begin{align}
\label{eq:AEMtaumugammamod}
   |A_\text{M}^{\tau\mu}|=& -\dfrac{1}{(4\pi)^2}gg^{SM}_{Z,A}\left( I_{\mu}^{+-} +  I_{\tau}^{+-} \right) ,  \nonumber \\
 |A_\text{E}^{\tau\mu}|=&\dfrac{i}{(4\pi)^2}gg^\text{SM}_{Z,V}\left(I_{\mu}^{--}+ I_{\tau}^{-+}\right) .
\end{align}
\section{V. Details on existing constraints in $Z'$ parameter space}
\label{App:Zpconstraint}
\begin{figure*}[t]
\centering
    \includegraphics[height=0.49\linewidth,width=0.49\linewidth]{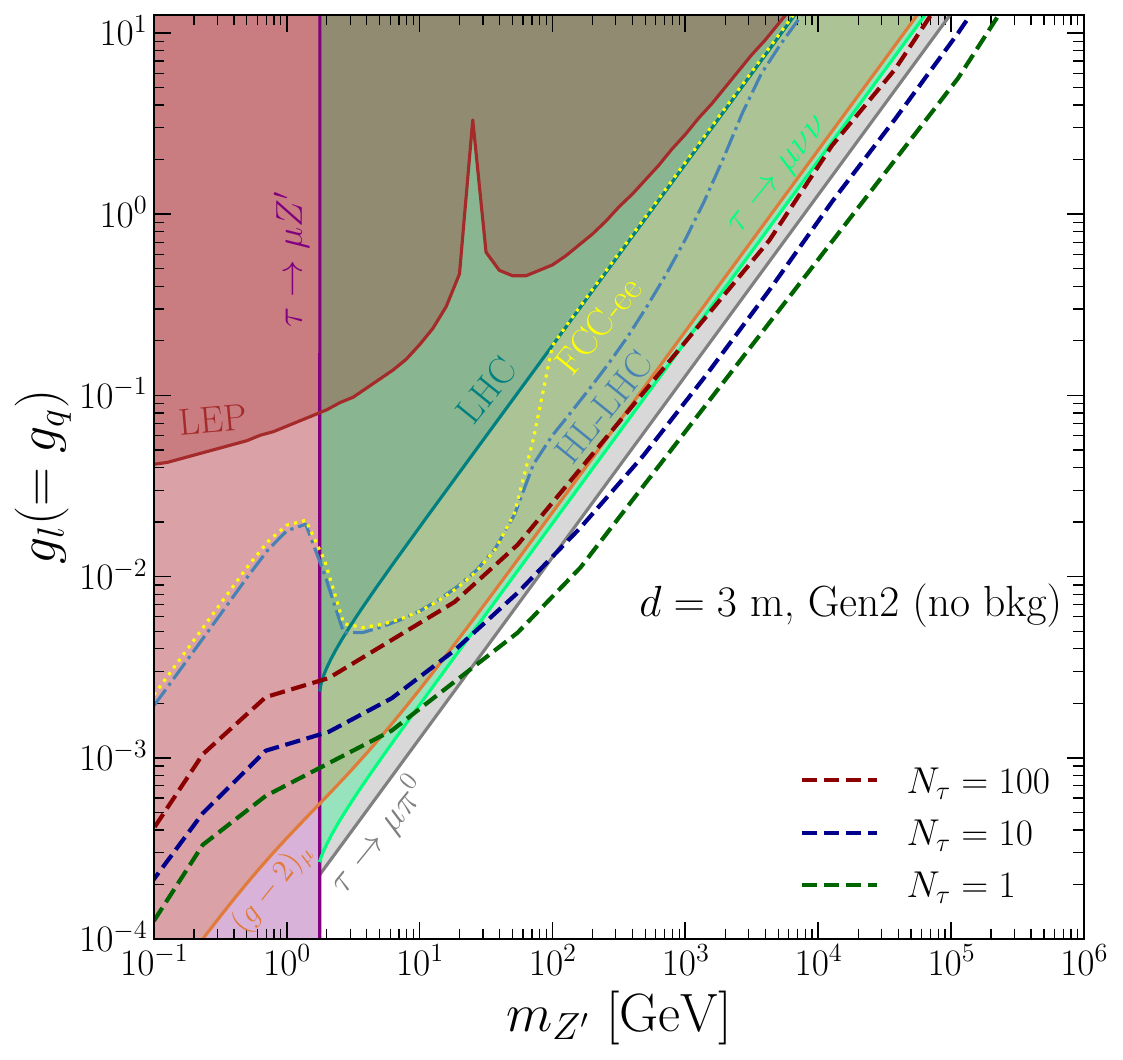}
    \includegraphics[height=0.49\linewidth,width=0.49\linewidth]{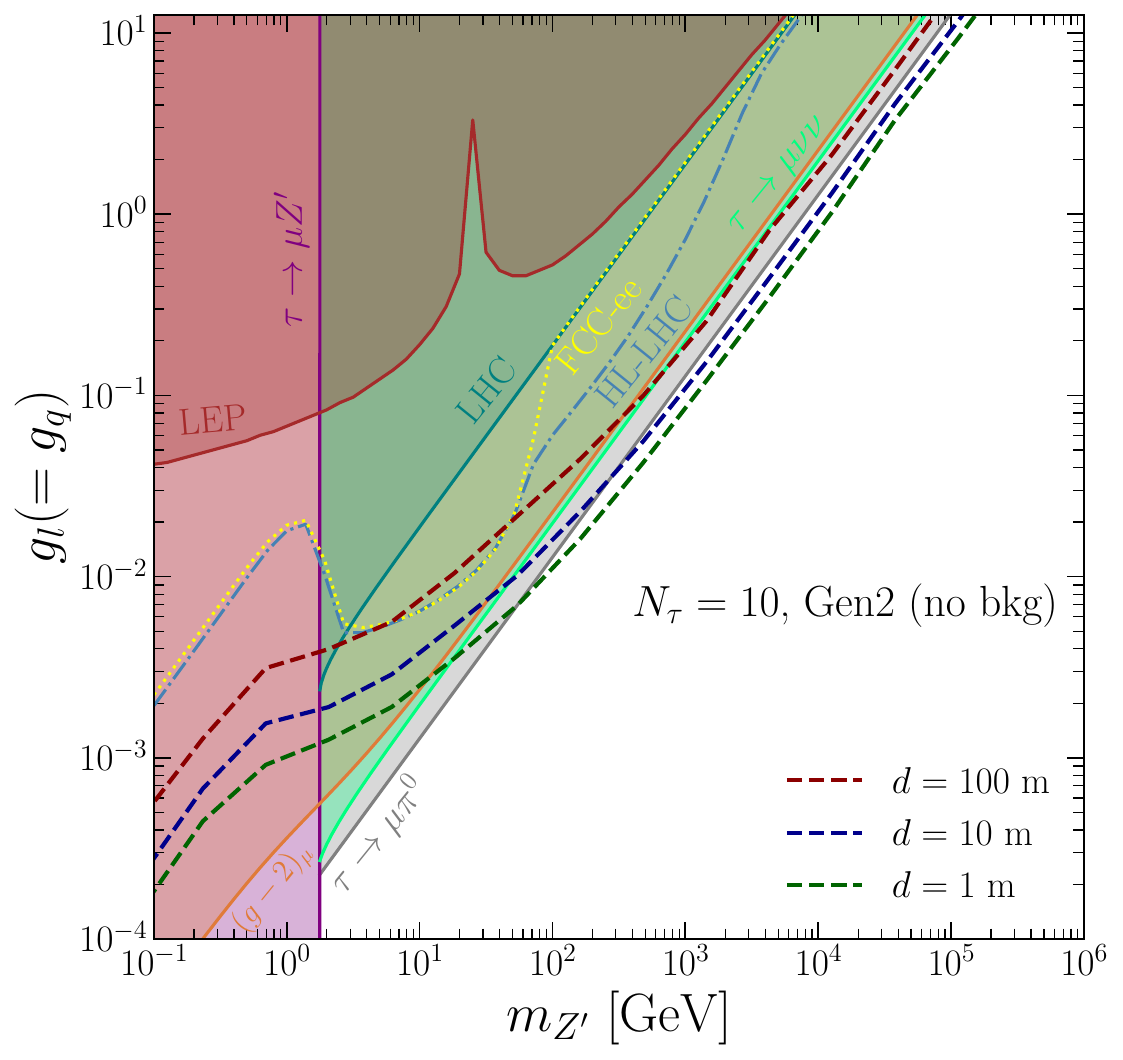}
    \captionsetup{justification=Justified}
    \caption{Sensitivity reach on the CLFV parameter space for an axial vector, $Z'$, projected for IceCube-Gen2 under a no-background assumption, for fixed tau survival distance and varying CLFV-induced event-number requirements (left panel) and for fixed event-number requirement and varying tau survival distances (right panel). The gray shaded parameter-space is excluded from the upper limit on  BR($\tau\rightarrow\mu\pi^0$) from BaBar~\cite{BaBar:2006jhm}. The green shaded region denotes the exclusion by the measurement of the ratio of $\text{BR}({\tau\rightarrow\mu\bar{\nu}_\mu\nu_\tau})$ to $\text{BR}({\tau\rightarrow e\bar{\nu}_e\nu_\tau})$ at Belle II~\cite{Belle-II:2024vvr} and the purple one corresponds to a similar measurement of $\tau$ decaying to muon plus an invisible particle at Belle II~\cite{Belle:2025bpu}. The salmon shaded parameter space is excluded by muon $g-2$ measurements~\cite{2:2026qau}. The teal and maroon shaded regions denote the exclusion of parameter space coming from the measurement of $pp\rightarrow W\rightarrow\tau\nu_\tau$ at LHC~\cite{ATLAS:2024irg} and from the precision measurement of $Z$ coupling at LEP~\cite{ALEPH:2005ab} respectively. Additionally, the steel blue dot-dashed curve and yellow dotted curve correspond to projected upper limit on $Z^\prime$ coupling from the sensitivity of HL-LHC and FCC respectively to a clean same-sign dilepton pair signal. For more details on these constraints, see  Supplemental Section V. Some of these constraints are updated from  Ref.~\cite{Altmannshofer:2016brv} taking into account the improved experimental measurements.}
    \label{fig:ZpFull}
\end{figure*}
The strongest constraint in our LFV $Z'$ parameter space comes from the search for LFV decay $\tau\to \mu\pi^0$. The Feynman diagram of this process is shown in Fig.~\ref{fig:FeynCLFV}(a). The decay width of this process takes the same form as mentioned in Eq.~\eqref{Eq:OpAxial} with suitable couplings and $\Lambda$ being replaced by $m_{Z^\prime}$.

The second strongest constraint comes from the search for $\tau\to \mu+{\rm invisible}$   at Belle II~\cite{Belle:2025bpu}. For $m_{Z'}<m_\tau-m_\mu$, our LFV $Z'$ given by Eq.~\eqref{eq:Zp} dominantly decays to $\nu\bar{\nu}$, hence the above search limit applies.  The decay-width of $\tau\rightarrow\mu Z^\prime$ looks like~\cite{Altmannshofer:2016brv}:
\begin{align}
\Gamma(\tau\to {\mu Z'}) &  = g^2 \frac{m_\tau^3}{16\pi m_{Z^\prime}^2} \Bigg[ \bigg\{ \left( 1 + \frac{2m_{Z^\prime}^2}{m_\tau^2} \right)\left( 1 - \frac{m_{Z^\prime}^2}{m_\tau^2} \right) \nonumber   - \frac{m_\mu^2}{m_\tau^2} \left( 2 - \frac{m_{Z^\prime}^2}{m_\tau^2} -\frac{m_\mu^2}{m_\tau^2}\right) \bigg\} + 6\,  \frac{m_\mu}{m_\tau} \frac{m_{Z^\prime}^2}{m_\tau^2} \Bigg] \nonumber \\
& \qquad \times \sqrt{\left( 1 - \frac{m_{Z^\prime}^2}{m_\tau^2} \right)^2 - \frac{m_\mu^2}{m_\tau^2} \left( 2 + \frac{2m_{Z^\prime}^2}{m_\tau^2} -\frac{m_\mu^2}{m_\tau^2}\right)  }\,.
\label{eq:muzp}
\end{align}

\begin{figure}[t]
\centering
\includegraphics[width=0.99\textwidth]{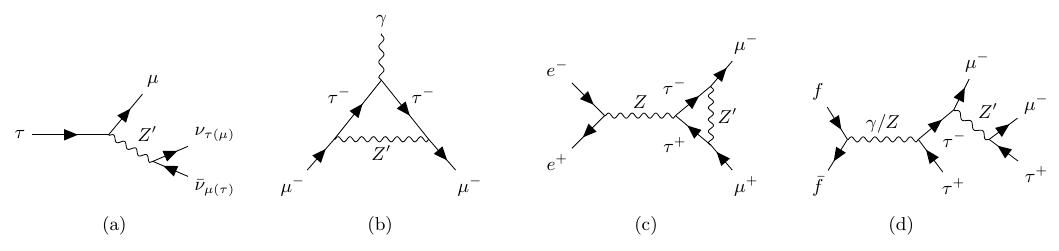}
\captionsetup{justification=Justified}
\caption{Feynman diagrams for (a) $Z^\prime$ mediated  decay of $\tau$ to muon and neutrinos, (b)  $Z^\prime$ affecting the anomalous magnetic moment of muon, (c) $Z'$-induced correction to $e^+e^-\to \mu^+\mu^-$ (and similarly to $\tau^+\tau^-$), and (d) $pp$ or $e^--e^+$ collision producing $\mu^\pm\mu^\pm\tau^\mp\tau^\mp$ signal via $Z'$.}
\label{fig:FeynZp}
\end{figure}

There are various other constraints that exist within the parameter space excluded by the above-mentioned two strongest constraints. One such constraint again comes from Belle II~\cite{Belle-II:2024vvr} in the measurement of the ratio of BR($\tau \rightarrow \mu \nu \bar{\nu}$) to BR($\mu \rightarrow e \nu \bar{\nu}$):
\begin{equation}
    \mathcal{R}\equiv\dfrac{\text{BR}(\tau \rightarrow \mu \nu \bar{\nu})}{\text{BR}(\mu \rightarrow e \nu \bar{\nu})}=0.9675\pm0.0007\pm0.0036.
\end{equation}
This measured $\mathcal{R}$ is in tension with the SM predicted value~\cite{Pich:2013lsa} at $2\sigma$ level. Existence of $Z^\prime$ with $\mu\tau$ LFV can affect the numerator of the quantity $\mathcal{R}$ (see  Fig.~\ref{fig:FeynZp}(a)). In the presence of such CLFV interaction, we get 
\begin{eqnarray}
\frac{R}{R^\text{SM}} & \ = \ & 1 + \frac{g^2}{g_2^2} \frac{4 m_W^2}{m_{Z^\prime}^2}f\left(\frac{m_\tau^2}{m_{Z^\prime}^2}\right)+ \nonumber + \dfrac{g^4}{g_2^4} \frac{4 m_W^4}{m_{Z^\prime}^4} g\left(\frac{m_\tau^2}{m_{Z^\prime}^2}\right) ,
\label{eq:genZp}
\end{eqnarray}
with $g_2$ being the the $SU(2)_{L}$ gauge coupling in the SM, and the functions
\begin{eqnarray}
 f(z) &=& \frac{2}{z^4} \left[ \frac{5}{6} z^3 + 2z^2 -2z - (1-z)^2(2+z)\log(1-z) \right] \,, \nonumber \\
 g(z) &=& \frac{2}{z^4} \left[ - z^3 -3z^2 +6z + 6(1-z)\log(1-z) \right] \,. \nonumber
\end{eqnarray}
Comparing this to the measured value of $\mathcal{R}$ and the SM prediction leads to a constraint shown by the green colored curve in the $Z^\prime$ parameter space (Fig.~\ref{fig:ZpFull}).

The existence of $Z^\prime\mu\tau$ vertex can also affect the anomalous magnetic moment of muon (see  Fig.~\ref{fig:FeynZp}(a)) in the following way:
\begin{align}
a_\mu   =  & \frac{m_\mu^2}{4\pi^2}\int_0^1 dx\Bigg[g^2\bigg\{(x-x^2)\left(x-\frac{2m_\tau}{m_\mu}-2\right)  -\frac{x^2}{2m_{Z'}^2}(m_\tau+m_\mu)^2\left(x+\frac{m_\tau}{m_\mu}-1\right)\bigg\}\Bigg]\nonumber \\ & \times \Big[m_\mu^2 x^2+m_{Z'}^2(1-x)+x(m_\tau^2-m_\mu^2)\Big]^{-1} \, .
\label{gm2}
\end{align}
Since the latest $g-2$ measurement at Fermilab~\cite{2:2026qau} is consistent with the updated SM prediction~\cite{Aliberti:2025beg}, it sets a constraint on the  $Z^\prime$ parameter space, as shown by the salmon curve in Fig.~\ref{fig:ZpFull}.

The measurement of the leptonic decay of $W$ boson at the LHC also sets a constraint on the $Z^\prime$ parameter space as $pp\rightarrow W\rightarrow \mu\nu$ will have a contribution from $pp\rightarrow W\rightarrow \tau\nu$, followed by the $Z^\prime$ mediated tau-decay:
\begin{eqnarray}
 \sigma(pp\to W\to \tau\nu_\tau \to \mu\nu_\tau\nu_\mu\nu_\tau) \ =  \ \sigma_{\rm SM}(pp\to W\to \tau\nu_\tau)   \times\:  {\rm BR}(\tau \to \mu\nu_\tau\nu_\mu\nu_\tau)\, .
\end{eqnarray}
For the SM predicted cross-section of $pp\to W\to \tau\nu_\tau$ and  $pp\to W\to \mu\nu_\mu$, we have used the NNLO predictions based on the CT14NNLO PDF set~\cite{ATLAS:2024irg} and compared the difference between the measured $pp\rightarrow W\rightarrow \mu\nu$~\cite{ATLAS:2024irg} and the SM-predicted cross-section  to get the teal colored exclusion curve in Fig.~\ref{fig:ZpFull}.

Similarly, the precise measurement of SM $Z$ coupling to all the leptons at LEP~\cite{ALEPH:2005ab} sets a constraint on the $Z^\prime$ parameter space (maroon curve in Fig. \ref{fig:ZpFull}). The effect of $Z'$ on $Z$-lepton couplings takes the following form (see Fig.~\ref{fig:FeynZp}(c)):
\begin{eqnarray}
\frac{g_{L\tau}}{g_{Le}} \ \simeq \ \frac{g_{L\mu}}{g_{Le}} & \ \simeq \ & 1 + \frac{g^2}{16\pi^2} ~\mathcal{K}(m_Z^2/m_{Z^\prime}^2) \, , \\ 
\frac{g_{R\tau}}{g_{Re}} \ \simeq \ \frac{g_{R\mu}}{g_{Re}} & \ \simeq \ & 1 + \frac{g^2}{16\pi^2} ~\mathcal{K}(m_Z^2/m_{Z^\prime}^2) \, , \\ 
\frac{g_{L\nu}}{g_{Re}- g_{Le}} & \ \simeq \ & 1 + \frac{2}{3}\frac{g^2}{16\pi^2} ~\mathcal{K}(m_Z^2/m_{Z^\prime}^2) \, ,
\end{eqnarray}
with the loop function given by 
\begin{eqnarray}
 \mathcal{K}(x) &=& - \frac{4+7x}{2x} + \frac{2+3x}{x}\log x  
 - \frac{2(1+x)^2}{x^2}\big[\log x \log(1+x) + \text{Li}_2(-x) \big] \, ,
\label{loop}
\end{eqnarray}
where $\text{Li}_2(x) = - \int_{0}^x dt \log(1-t)/t$ is the di-logarithm.
In the above expressions, the electron couplings $g_{Le}$ and $g_{Re}$ are used as their measured values as these couplings are not affected by $Z'$ loops. 

Finally, the LFV $Z^\prime$ coupling gives rise to a clean signature of same-signed dilepton pair production at both $pp$ and lepton colliders (see Fig.~\ref{fig:FeynZp}(d)). We have performed a simple estimate of the signal sensitivity as $\mathcal{N} = S/\sqrt{S + B} \simeq \sqrt{\mathcal{L}\sigma_\text{signal}}$, where $\mathcal{L}$ is the integrated luminosity and $\sigma_\text{signal}$ is the signal cross section times efficiency, with cross-section obtained from \texttt{MadGraph}~\cite{Alwall:2014hca} simulation. A $70\%$ tau-tagging efficiency is assumed. The integrated luminosity is taken to be $3~\text{ab}^{-1}$ for HL-LHC~\cite{CidVidal:2018eel} and $2.6~\text{ab}^{-1}$ for FCC-ee~\cite{FCC:2018evy}.

All these constraints are obtained following the analysis of Ref.~\cite{Altmannshofer:2016brv}, but some of them have been updated using improved experimental results. It is important to note that Ref.~\cite{Altmannshofer:2016brv} considered $Z'$ to be a vector boson which has left-handed and right-handed couplings $g_L$ and $g_R$ respectively. In the limit of $g_R$ being equal to $-g_L$, we obtain the axial vector $Z^\prime$ mediated interactions considered here.

The two figures in Fig.~\ref{fig:ZpFull} correspond to the upper limit on $Z'$ coupling to leptons and quarks (assumed to be equal in our work) projected for IceCube-Gen2 under two different conditions. In the left panel, the survival distance of taus is chosen to be $3m$ and the requirement for the CLFV-induced events is varied. In the right panel, ten CLFV-induced tau events are demanded with varying tau survival distances. In both cases, we find that Gen2 can cover unconstrained $Z'$ parameter space using the new search topology proposed in this work.

\end{document}